# Thermal boundary conductance in standalone and non-standalone GaN/AlN heterostructures predicted using machine learning interatomic potentials

Hao Zhou[1], Khalid Zobaid Adnan[1], Wyatt Allen Jones[1], Tianli Feng[1]*

[1]Department of Mechanical Engineering, University of Utah, Salt Lake City, Utah 84112, USA

**Corresponding Authors**

*Email: tianli.feng@utah.edu

## Abstract

GaN/AlN interfaces are essential in advanced high-power and high-frequency electronic devices, where effective thermal management is crucial for optimal performance and reliability. This work investigates the thermal boundary conductance (TBC) of standalone and non-standalone GaN/AlN heterostructures using non-equilibrium molecular dynamics (NEMD) driven by accurate machine learning interatomic potentials trained from density-functional theory calculations. For the standalone interface, the TBC is found to be ~600 MW m$^{-2}$ K$^{-1}$ at room temperature after quantum correction, with possible ~60% elastic and ~40% inelastic contributions. The result revises previous NEMD predictions (400 – 2000 MW m$^{-2}$ K$^{-1}$) using empirical interatomic potentials. When a second GaN/AlN interface is brought close to the original interface, the TBC of the original interface can increase to 1000 MW m$^{-2}$ K$^{-1}$, and this value gradually decreases with increasing distance between the two interfaces. When multiple interfaces are introduced in proximity, the first interface's TBC can be further enhanced to above 1150 MW m$^{-2}$ K$^{-1}$. After comparing double interfaces, superlattices, and random multilayers, it is concluded that such enhancement of TBC is not caused by the emergence of coherent superlattice modes but rather by the ballistic transport of the original bulk phonon modes of each material. Additionally, a "critical separation distance" ($l_{cs}$) is defined as the threshold beyond which the two interfaces no longer influence each other and behave independently. $l_{cs}$ is determined by the mean free path of the phonons filtered by the interfaces that transport in the middle layer between two interfaces. Our findings provide insights into the thermal transport mechanisms that may aid the design of future electronic devices.

Keywords: GaN, AlN, Wide Band Gap Semiconductors, Interfacial Thermal Resistance, Molecular Dynamics

## I. INTRODUCTION

Gallium nitride, a III-V compound semiconductor, possesses exceptional properties, including a wide band gap, high electronic saturation velocity, high thermal conductivity, and a high critical electric field [1–3]. These attributes make GaN highly promising for high-power and high-frequency applications, such as high electron mobility transistors (HEMTs), power electronics, and optoelectronic devices [4–7]. As these devices are miniaturized and the power density is increased, Joule heating leads to elevated temperatures, which, if not efficiently dissipated, can severely degrade both reliability and performance [6]. A significant source of thermal resistance in these systems arises from the interfaces of semiconductor heterostructures, commonly characterized by thermal boundary resistance, or inversely, thermal boundary conductance (TBC, *G*) [8–10]. In GaN-based devices, GaN/AlN heterostructures are frequently present due to AlN's minimal lattice constant mismatch with GaN, allowing it to serve as a substrate or buffer layer with small strain at the interface [11–13]. Therefore, the TBC of GaN/AlN heterostructures plays a critical role in determining heat dissipation efficiency.

Despite its importance, the TBC of GaN/AlN heterostructures remains unclear. One key challenge is the lack of clarity around the intrinsic TBC of a standalone GaN/AlN interface. To date, only limited experimental measurements are available. Li et al. reported a TBC of approximately 320 MW $m^{-2}$ $K^{-1}$ using time-domain thermoreflectance (TDTR) [14]. More recently, Walwil *et al.* measured the TBC of the a standalone GaN/AlN interface to be around 490 MW $m^{-2}$ $K^{-1}$ [15]. In both cases, sample quality factors such as interfacial mixing, misfit, or dislocation may be present. Other experimental efforts focus on either superlattices, where nearby interfaces can affect the intrinsic TBC value, as discussed later, or the total thermal conductivity of the whole heterostructures [16,17]. For instance, Koh *et al.* extracted the TBC of one individual interface from a superlattice to be 620 MW $m^{-2}$ $K^{-1}$ at room temperature using TDTR in 2009 [16]. While experimental data remain scarce, many simulation studies have been conducted [18–27]. The most commonly used method is non-equilibrium molecular dynamics (NEMD), but the predicted TBC values vary significantly, ranging from 200 to 3000 MW $m^{-2}$ $K^{-1}$ at room temperature (without quantum correction), due in part to the unreliable empirical interatomic potentials. Additionally, quantum corrections are frequently neglected in these simulations, leading to significant errors, as NEMD relies on classical statistics. Landauer formula [20] and Boltzmann transport equation (BTE) [26] have also been used to study the TBC of GaN/AlN interface. However, they rely on the implementation of mode-dependent interfacial transmission coefficients, which are hard to predict accurately using existing models, such as the acoustic mismatch model [28,29], diffuse mismatch model (DMM) [30], and the higher harmonic inelastic model [31]. Other approaches, such as the non-equilibrium Green's function (NEGF) approach, predict a TBC of around 300 MW $m^{-2}$ $K^{-1}$, which ignores the anharmonic effects, potentially limiting its accuracy [14,20,24].

Additionally, devices often contain more than one closely spaced interface, such as Fig. 1(a), raising critical questions: How much does the TBC of an individual interface depend on a neighboring interface? Under what conditions—specifically, at what interfacial spacing—does this influence become significant? Is this multi-interface coupling effect exclusive to superlattices with numerous interfaces, or does it also occur when only two interfaces are present? In the literature, the coupling between multiple interfaces has been extensively studied in superlattices, where the coupling was often attributed to the emergent periodicity of the superlattice junction [32–41]. The periodicity was claimed to induce the wave nature of phonons, which undergo coherent reflections at the equally spaced interfaces. These reflections lead to the formation of mini-bands and new eigenmodes known as superlattice phonon modes [37]. These new phonon modes cannot "see" the interfaces, thereby traversing the interfaces freely, which enhances the TBC of each individual interface, as illustrated in Fig. 1(b). This coherent effect is found to vanish once the superlattice period is randomized, disrupting coherence, as shown in Fig. 1(c).

Such enhancement in TBC has also been reported in systems with only a few closely spaced interfaces, in the absence of periodicity or even with only two interfaces [42–45]. Landry and McGaughey studied the TBC of Si/Ge/Si and Ge/Si/Ge junctions using both molecular dynamics and lattice dynamics (scattering boundary method) [44]. They found that the total resistance of the middle structure—including the two interfaces and the middle thin film—decreases with decreasing film thickness. The reason is attributed to the reduced phonon-phonon scattering in the thin film as the film thickness becomes smaller than the phonon mean free path. Such reduced inelastic scattering increases the coupling of phonons at the two interfaces and decrease the thermal resistance of the film. They estimated that this effect vanishes entirely when the film thickness exceeds 10 μm. Recently, Adnan and Feng explicitly studied the TBC of each individual interface in two-interface systems [45]. The pointed out that TBCs of nearby interfaces influence each other through the ballistic effects of the original bulk phonon modes and the selective transmission of phonon of each interface, rather than the coherent effect of newly formed superlattice modes. Specifically, when the equilibrium bulk phonon modes reach an interface, only a subset – termed "selected phonon modes" – is transmitted. If the interface is in close proximity to another interface, these selected phonon modes can travel ballistically to the second interface, meaning that the phonons incident on the second interface are already filtered by the first, as illustrated in Fig. 1 (d). In contrast, when the spacing between the two interfaces is sufficiently large, this effect vanishes because the selected phonons have enough time and space to re-equilibrate with other phonon modes via phonon-phonon scattering, ensuring that the phonons incident on the second interface are again in equilibrium with the bulk, as shown in Fig. 1 (e). Therefore, the incident phonon distribution at the second interface depends on its separation from the first, directly impacting its TBC. Since heat transport is reciprocal [46], the second interface affects the first in the same manner. Adnan and Feng demonstrated this effect in "toy" Si/Ge heterostructures using empirical interatomic potentials self-consistently. To date, however, this effect has yet to be explored for more realistic GaN/AlN interfaces using more accurate interatomic potentials.

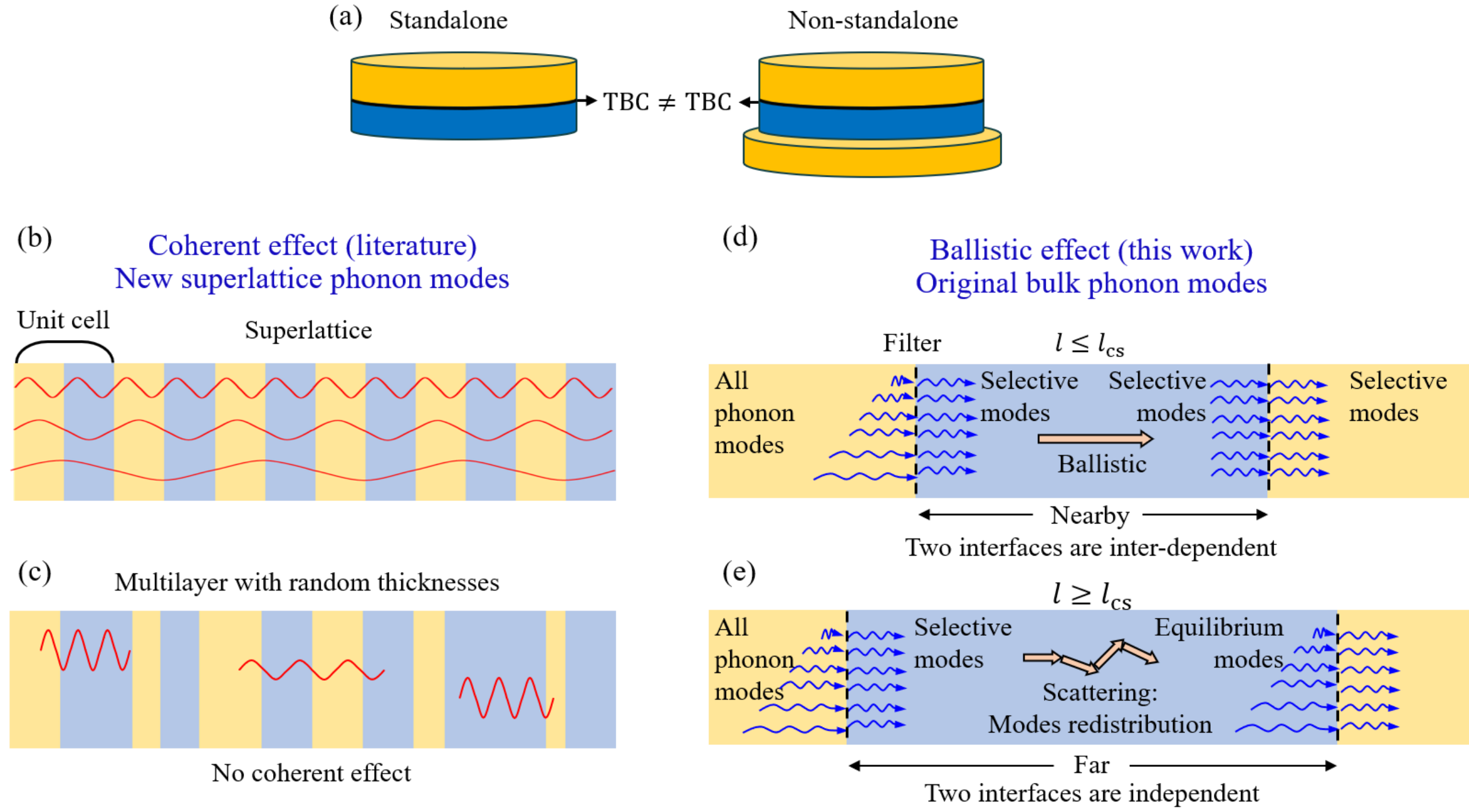

FIG. 1. Schematic diagrams for standalone and non-standalone interfacial thermal transport. (a) A standalone vs. a non-standalone interface. (b) The coherent effect. A periodic structure can create emerging superlattice phonon modes, with wavelength being integer times of periodic length, contributing to the overall transport. (c) Random multilayers break the coherence and eliminate the emerging superlattice phonon modes. (d) The ballistic effect. The phonon modes that pass through one interface reach the second interface ballistically due to the short distance between the two interfaces. (e) The phonon modes that pass through one interface get equilibrated with other modes before reaching the second interface due to the long distance between the two interfaces. Note that although the sketch depicts a one-way impact, the interaction is actually bidirectional.

In this work, we investigate the TBC of GaN/AlN heterostructures using NEMD with a machine learning interatomic potential (MLIP). The recently emerged MLIP has been demonstrated to achieve first principles accuracy for various systems [49–53]. The quantum-corrected theoretical TBC for both standalone and non-standalone GaN/AlN interfaces are calculated, along with its size effects and temperature dependence. The impact of nearby interfaces on TBC values is investigated for double-interface and multilayer structures. The underlying reason for the coupling between nearby interfaces is elucidated.

## II. METHODOLOGY

The GaN/AlN interface is constructed by aligning wurtzite hexagonal GaN and AlN in the [001] direction ($z$-axis), as shown in the inset of Fig. 3, along which the lattice mismatch strain is minimal [11–13]. The interface is relaxed using an iterative trial-and-error approach. Specifically, *ab initio* molecular dynamics (AIMD) simulations are conducted in the canonical ensemble (NVT), during which the stresses along the three cartesian directions and lattice parameters are recorded. Based on Young's modulus, the lattice parameters are manually adjusted, and the process is repeated until the stresses along all directions approach zero. After that, four independent AIMD simulations under the NVT ensemble are run at each temperature (100, 300, 450, and 600 K) with distinct initial velocities to sample the potential energy surface. The time step of AIMD is 1 fs, and the total length of each AIMD simulation is 1.5 ns. During AIMD simulations, the atomic energies, forces, and stresses of each snapshot are recorded to construct the MLIP database. All the first-principles calculations use density function theory (DFT) implemented by the Vienna ab initio simulation package (VASP) [54] with the projector augmented wave (PAW) [55] method and the local-density approximation (LDA) exchange-correlation functional [56,57]. The electron energy convergence threshold is selected as $2\times10^{-6}$ eV and the plane-wave energy cutoff is 400 eV, with a $\Gamma$ only **k**-point for electron. Periodic boundary conditions are applied for all three cartesian directions.

The Moment Tensor Potential (MTP) developed by Novikov et al. and implemented in the MLIP package is selected in this work for its simplicity and computational efficiency [58,59]. The potential is trained over 1000 iteration steps with the minimum and maximum atomic interaction cutoffs of 1.2 Å and 4.5 Å, respectively. The minimum cutoff is smaller than the minimum atomic distance in both the training and testing databases. The maximum cutoff is large enough to include the dominant interatomic interactions, and smaller than half of the AIMD simulation dimension to avoid the periodic images' interaction. The accuracy of MTP is examined by comparing the predicted energies and forces to those in the DFT testing database. As demonstrated in Fig. 2(a,b), the root-mean-square errors (RMSEs) for both energies and forces are small. To further test the accuracy, the MLIP is used to calculate the phonon dispersions of bulk GaN and AlN and is compared with DFT results. Despite being trained solely on the GaN/AlN heterostructure database, which has different lattice constants from bulk GaN and AlN, the MLIP surprisingly predicts the phonon dispersions of bulk GaN and AlN with remarkable accuracy, as shown in Fig. 2(c,d). Its performance is significantly superior to that of the Tersoff and Stillinger-Weber (SW) potentials [21] [60,61].

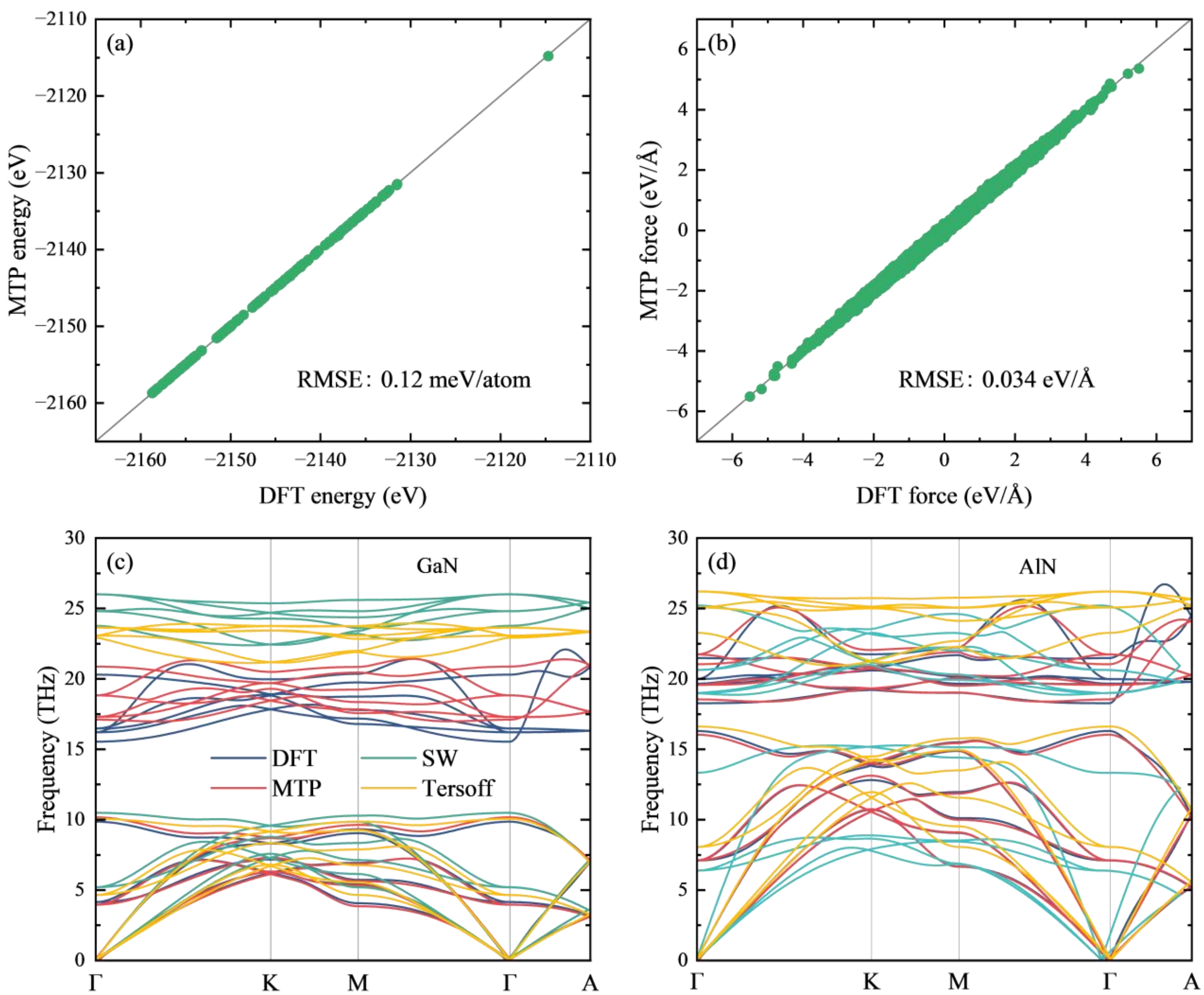

FIG. 2. Validation of the trained MTP for the GaN/AlN interface. (a, b) Comparison of energies and forces predicted by MTP versus DFT results from the testing database. (c, d) Phonon dispersions of GaN and AlN computed using MTP, SW, and Tersoff interatomic potentials, benchmarked against DFT calculations.

The TBCs of GaN/AlN heterostructures are calculated by using NEMD based on the large-scale atomic molecular massively parallel simulator (LAMMPS) package [62]. For the NEMD setup, the post-AIMD structure is extended along the $z$ axis by replicating the middle bulk region of each material, leaving the interface regions unchanged. Periodic boundary conditions are applied along the x- and y-axes. The two edges along the $z$ direction, containing 64 atoms on each edge, are fixed, to mimic an adiabatic boundary condition. Next to the fixed edges, the Langevin thermostat is applied to a 4 nm region on each side. The systems are initially equilibrated under NVT ensemble for 1 ns at 300 K. Then the temperatures of two reservoirs are changed to 330 and 270 K using the Langevin thermostat to introduce temperature gradient in the systems. After that, the system is run with the microcanonical ensemble (NVE) for 2 ns to establish the temperature profile. Subsequently, the system is run for another 2 ns under NVE to collect the temperature and heat flux ($Q$) data. For large systems, the simulations are performed with a longer time to ensure a steady state temperature profile. The TBC value is obtained using $G_{\mathrm{MD}} = \frac{Q}{A\Delta T}$, where $A$ is the cross-sectional area and $\Delta T$ is the temperature jump at the interface. Each simulation is repeated at least three times with different initial velocity assignments, and the error bars represent the standard error across these simulations. As MD simulations inherently follow classical statistics, a quantum correction is applied to the obtained TBC using $G = G_{\mathrm{MD}} \frac{c_Q}{c_C}$, where $c_{\mathrm{Q}}$ and $c_{\mathrm{c}}$ are the quantum and classical specific heats of GaN at the studied temperature. At 300 K, $\frac{c_{\mathrm{Q}}}{c_{\mathrm{C}}} \approx 0.7134$. GaN is chosen over AlN for specific heat correction because GaN's phonon modes have lower energies than AlN and thus dominate the thermal transport across the interface [63]. Nonetheless, selecting GaN or AlN has a small impact on the final results.

## III. STANDALONE INTERFACE

The obtained room-temperature TBC (after quantum correction) of the standalone GaN/AlN interface as a function of thickness is shown in Fig. 3. The converged TBC value is about 600 MW $m^{-2}$ $K^{-1}$ at around 21 nm. The domain size effect is small, compared to those by classical potentials-driven MD simulations [14,18,19,64,65]. This can be partially attributed to the use of Langevin thermostat and MLIPs, as seen in other systems as well [63,66–69]. The 21-nm length is selected to study the temperature dependence.

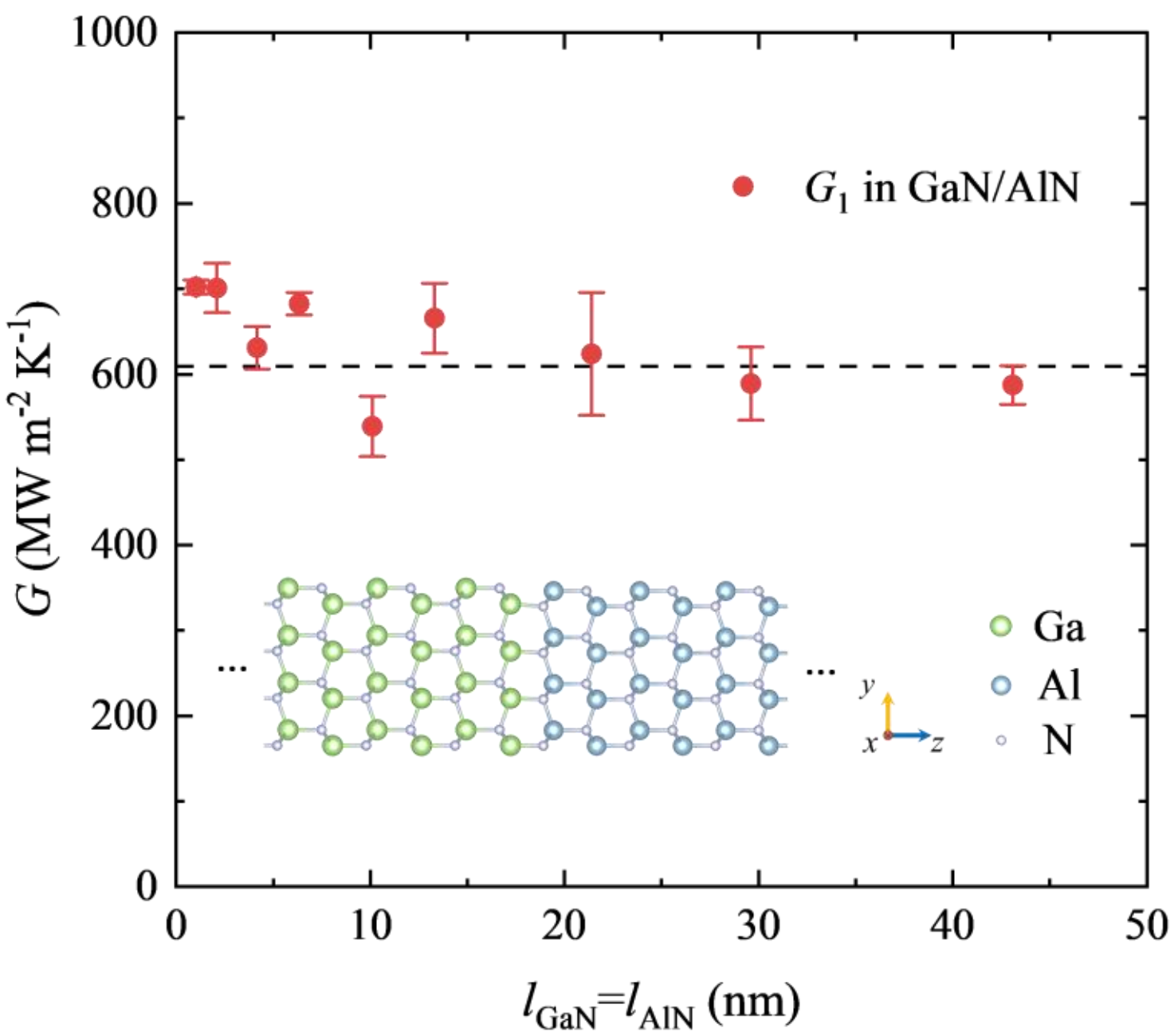

FIG. 3. Size effect of TBC of a standalone GaN/AlN interface simulated using MLIP-NEMD.

The temperature-dependent TBC of a standalone GaN/AlN interface is shown in Fig. 4. The TBC predicted by MLIP-NEMD rises with temperature and saturates at approximately 630 MW m$^{-2}$ K$^{-1}$ above 400 K, near the Debye temperature of GaN (600 K) [70]. These TBC values are much higher than experimental data (320 and 490 MW m$^{-2}$ K$^{-1}$) [14,15], which can be attributed to two possible reasons. First, achieving atomically sharp interfaces in experiments is challenging. Typically, a 2-nm or thicker amorphous or oxide layer often develops at the interface [71–73]. In addition, atomic-scale interfacial mixing may be present in experimental samples, which can reduce TBC by increasing phonon reflectance [74]. Second, experimental samples may have atomic misfits at the interface due to the lattice mismatch between GaN and AlN, while in the simulations the two lattices are enforced to match at the interface [21]. It is worth noting that the TBC reported by Walwil *et al.* shows an abnormal temperature dependence, decreasing with increasing temperature. The authors attribute this behavior to the high defect density near the interface. This may also explain the large error bars associated with the measurements.

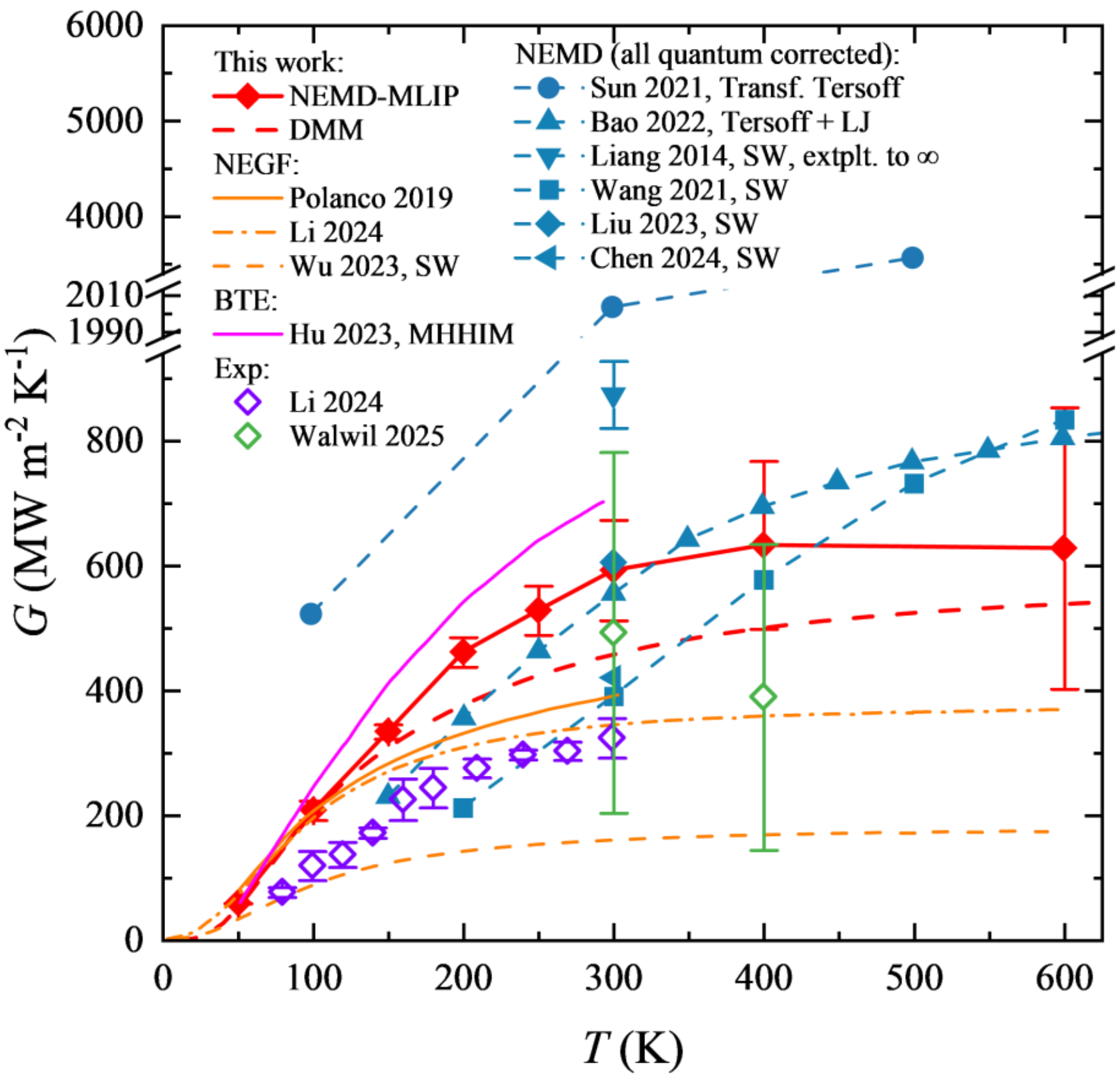

FIG. 4. Temperature-dependent TBC of a standalone GaN/AlN interface simulated using MLIP-NEMD, compared to the DMM and other methods in the literature. All MD simulated results are quantum corrected using specific heat. References: Sun 2021 [21], Bao 2022 [23], Liang 2014 [18,19], Wang 2021 [22], Liu 2023 [25], Chen 2024 [27], Polanco 2019 [20], Li 2024 [14], Wu 2023 [24], Hu 2023 [26].

Our results are compared to several other theoretical predictions in the literature, as shown in Fig. 4. The most common approach is NEMD with empirical interatomic potentials [18,19,21–23,25,27]. Unlike MLIP, empirical potentials often fail to accurately reproduce atomic interactions, leading to a wide range of predicted TBC values. As can be seen from Fig. 2(c,d), empirical potentials predict less frequency mismatch between GaN and AlN than MTP. Therefore, the predicted TBC using empirical potentials are generally higher than MTP. For example, transferable Tersoff potential by Sun *et al*. predicted a TBC of over 2000 MW $m^{-2}$ $K^{-1}$ at room temperature after quantum correction [21]. Bao *et al*. used the Tersoff potential to model GaN and AlN and the Lennard-Jones (LJ) potential for interfacial interaction [23]. They obtained 550 MW $m^{-2}$ $K^{-1}$ at 300 K, which converges to 800 MW $m^{-2}$ $K^{-1}$ at high temperature. The others used SW potential and obtained TBC values ranging from 390 to 873 MW $m^{-2}K^{-1}$ at 300 K [18,19,22,25,27] due to the use of different system sizes. Another reason that empirical potentials cannot get reasonable TBC values is that empirical potentials usually have only been fitted to the bulk material, but not validated on the interfacial interaction [75]. In summary, none of the empirical potentials-based NEMD simulations yield similar TBC as our MLIP-based simulations. Note that we have applied quantum corrections to all the classical TBC data obtained in MD simulations in the literature.

NEGF is another popular method to calculate TBC. Among the three available NEGF results, Wu *et al*. employed an empirical interatomic potential, resulting in TBC values that significantly differ from the other two first principles-based simulations. This discrepancy further highlights the limitations of empirical potentials [24]. Noteworthy, the two first principles-based NEGF calculations [14,20] produce TBC values similar to our MLIP-based NEMD results at low temperatures but show significant deviations above 100 K. This might indicate that elastic transmission dominates the TBC at low temperature, while as temperature increases, inelastic transmission begins to play a role, which is not included in NEGF. This explains the deviation between NEGF and MLIP-based NEMD at higher temperatures. The discrepancy between the two methods suggests that inelastic transmission contributes ~40% to the TBC at room temperature, highlighting its critical role in thermal transport across the interface. Likewise, both the DMM and BTE using DMM's transmission coefficient [26] account only for elastic transmission, resulting in a temperature dependence similar to that of NEGF. However, DMM's transmission coefficients are significantly less accurate than those from NEGF, since DMM does not consider the bonding strength between two materials.

The TBC of the GaN/AlN interface is notably higher than those of many other interfaces predicted by MLIPs, such as Si/diamond (122 MW $m^{-2}$ $K^{-1}$) [67], diamond/metal (40–300 MW $m^{-2}$ $K^{-1}$) [63], and GaN/SiC (529 MW $m^{-2}$ $K^{-1}$) [66]. It is also higher than Al/diamond, GaN/SiC, and GaN/diamond, whose TBC values typically fall below 300 MW $m^{-2}$ $K^{-1}$ based on experiments [66,76–78]. The high TBC value underscores the strong thermal transport capability of GaN/AlN heterostructures in comparison to other systems. High thermal conductivity in the individual materials does not necessarily translate to high TBC at their interfaces. For instance, although SiC [79], diamond [80], and metals [81] have higher intrinsic thermal conductivities than GaN [1] and AlN [82,83], their interface TBCs are lower. Similar high TBC values are seen in interfaces with minimal lattice mismatch, such as ZnO/GaN (~500 MW $m^{-2}$ $K^{-1}$) [84] and $SrRuO_3/SrTiO_3$ (~800 MW $m^{-2}$ $K^{-1}$) [85], where the constituent materials have relatively low thermal conductivities. This points to the importance of structural compatibility at the interface.

## IV. DOUBLE INTERFACES

After obtaining the intrinsic TBC of a standalone GaN/AlN interface, we examine the impact of nearby interfaces. First, we investigate the impact of adding a second interface by creating a GaN/AlN/GaN double-interface system. The lengths of GaN layers are fixed at 4 nm for the sake of computational cost. This is why the $G_1$ in GaN/AlN is slightly higher than 600 MW $m^{-2}$ $K^{-1}$, as discussed in Fig. 3. The TBC of the first interface ($G_1$) is obtained as a function of the thicknesses of the intermediate AlN layer. As shown in Fig. 5, $G_1$ is significantly enhanced from around 600 to 1000 MW $m^{-2}$ $K^{-1}$ when the two interfaces' spacing is thinner than 10 nm. This enhancement gradually diminishes as the interfacial spacing increases, although at a separation of 80 nm, the TBC remains elevated by approximately 100 MW $m^{-2}$ $K^{-1}$.

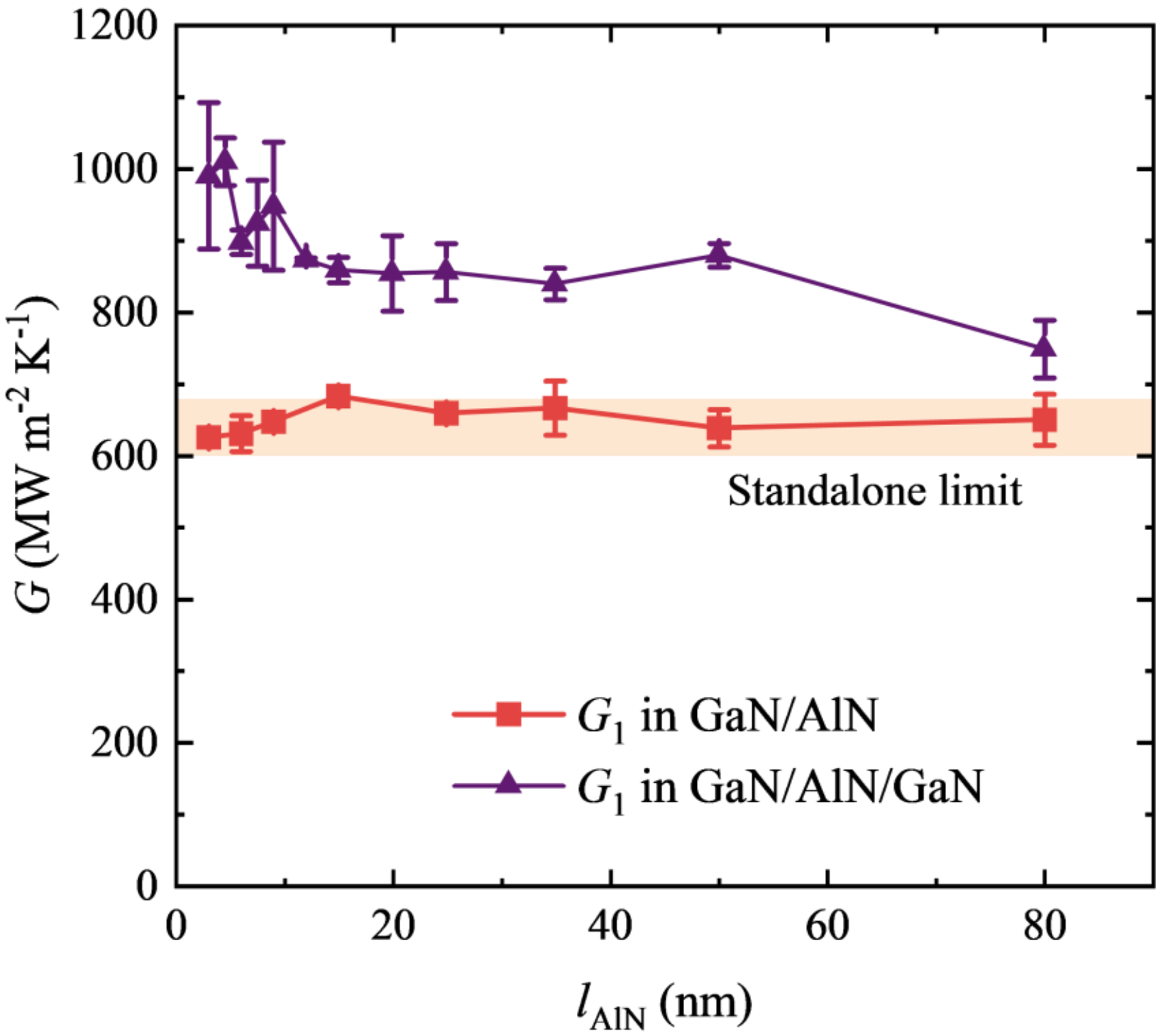

FIG. 5. TBC of a standalone and a non-standalone GaN/AlN interface obtained by MLIP-NEMD simulations after quantum correction as a function of AlN thickness. The thickness of each GaN layer is 4 nm for those simulations.

To understand the mechanism behind this enhancement, we first note that coherent wave effects are unlikely to be the cause, as there is no periodic structure in this double-interface system. Another plausible explanation is the formation of interfacial modes at the interface, arising from the structural differences between the interface and the bulk regions. These modes have been observed both experimentally and theoretically across various systems [86–88]. Interfacial modes can form even in single-interface systems and may serve as bridging states that facilitate the transmission of bulk phonon modes through the interface. A double-interface system could potentially amplify this effect, as the whole structure becomes more different from the bulk. However, Feng *et al*. demonstrated that such interfacial modes are typically confined to within approximately 1 nm of the interface, beyond which bulk modes dominate the transport [89]. Explicitly accounting for interfacial modes would introduce additional complexity into the analysis and falls outside the scope of the present study. Therefore, it is reasonable to treat the interface region as a black box and focus solely on the transport of bulk phonon modes.

The observed enhancement appears to originate from ballistic phonon transport. When the AlN layer is thin, phonons selectively pass through one interface, travel ballistically to the other, and fully transmit it since the two interfaces are identical. As the separation between two interfaces increases, the selected phonon modes scatter with other modes in AlN and start to become equilibrium, mitigating the enhancement. At sufficiently long separation, we expect that phonons recover their equilibrium distribution when they reach the second interface, making the TBC of the second interface independent of the first, in other words, the disappearance of the enhancement.

Based on this behavior, we introduce "critical separation distance" $l_{\mathrm{cs}}$, representing the threshold distance beyond which the impact of nearby interfaces becomes negligible and the interfaces become independent. $l_{\mathrm{cs}}$ should be longer than the mean free path (MFP) of the phonon modes that pass through the interfaces and transport inside the sandwiched material. For example, if the most transmitted phonons are acoustic phonons with long MFP, then $l_{\mathrm{cs}}$ would be long. If the most transmitted phonons are optical phonons, $l_{\mathrm{cs}}$ would be short. This might explain why the coupling between two interfaces was not observed in Ref. [90] since the interfaces are rough, which may allow short MFP phonons to transmit, resulting in a small $l_{\mathrm{cs}}$. Our previous work based on the Tersoff potential shows that $l_{\mathrm{cs}}$ of the Si/Ge/Si system is larger than 160 nm, longer than the average MFP of phonons in Ge (102 nm) [45], due to the perfect lattice match at the interface. The $l_{\mathrm{cs}}$ for the GaN/AlN/GaN system is hard to simulate due to the computational cost of MLIP-MD, and we expect it at least to be longer than the average MFP of AlN (300 nm [91]) since the lattices at the interface are well matched.

Some clues can be found in the literature about the magnitude of $l_{\mathrm{cs}}$ for AlN/GaN/AlN. The mean TBC values of each interface in $(\mathrm{AlN})_x/(\mathrm{GaN})_y/(\mathrm{AlN})_x$ tri-layers are measured by Koh *et al*. using TDTR in 2009, where thickness $x$ = 4 nm and $y$ ranges from 200 to 1000 nm [16]. They reported a mean TBC of approximately 620 MW m$^{-2}$ K$^{-1}$ per interface, independent of $y$. While these findings appear to suggest that the $l_{\mathrm{cs}}$ is smaller than 200 nm, it might not be true. In their paper, TBC was derived based on the thermal resistance model: $\frac{2}{G} = \frac{x+y}{\kappa} - \frac{x}{\kappa_{\mathrm{AlN}}^{\mathrm{Bulk}}} - \frac{y}{\kappa_{\mathrm{GaN}}^{\mathrm{Bulk}}}$, where $\kappa$ is the total thermal conductivity of the tri-layer structure, and $\kappa_{\mathrm{AlN}}^{\mathrm{Bulk}}$ and $\kappa_{\mathrm{GaN}}^{\mathrm{Bulk}}$ are the bulk thermal conductivity of AlN and GaN, respectively. Since the system is at the nanoscale, using bulk thermal conductivities to evaluate thin films would lead to an underestimation of TBC, especially for small thicknesses (small $y$). In other words, if the thermal conductivities are corrected to account for size effects, the TBC should decrease with increasing $y$, indicating that they had not yet reached the independent interface regime, i.e., $l_{cs}$ should be larger than 200 nm. Landry and McGaughey also estimated that fully diffusive phonon transport is not expected until the film thickness is around 10 μm for Si/Ge/Si and Ge/Si/Ge systems [44]. These evidences all indicates that $l_{cs}$ should be much larger than the average phonon MFP of the middle layer material.

The thermal conductivity of the intermediate AlN layer ($\kappa_{AlN}$) can also be influenced by the addition of nearby materials, as shown in Fig. 6. Note that the thermal conductivity data is compared in a self-consistent manner within the MD framework rather than against experimental data, due to the classical limitation of MD. Here, we consider the conventional thermal conductivity, defined as the heat flux divided by the linear-region temperature gradient in AlN [45]. A 50 nm-thick standalone AlN film has a thermal conductivity of 56 W $m^{-1}$ $K^{-1}$. After attaching to GaN on one side, its thermal conductivity increases to 70 W $m^{-1}$ $K^{-1}$. If both sides are attached to GaN, AlN's thermal conductivity increases to 84 W $m^{-1}$ $K^{-1}$. The increases can be understood. First, the higher atomic mass of GaN leads to lower-frequency phonons (acoustic phonons) in the terminal layers, which cross the interfaces and lead to more acoustic phonons in the AlN layer. As acoustic phonons generally contribute more to thermal conductivity, $\kappa_{AlN}$ is elevated. This aligns with our findings in the Si/Ge system, where the thermal conductivity enhancement of Si in a Ge/Si/Ge structure exceeds that of Ge in a Si/Ge/Si structure due to Ge's larger atomic mass [45]. Second, the presence of additional layers effectively increases the AlN's equivalent thickness, allowing some phonons—previously limited by the finite AlN size—to propagate with longer MFP, thereby raising $\kappa_{AlN}$. We expect this enhancement to vanish once the separation distance significantly exceeds the phonon MFP. This observation suggests that thermal conductivity cannot be assumed to match either the bulk value or the value at that length scale when a material is embedded in a multilayer structure, such as in HEMTs. This enhancement is not observed for AlN thicknesses below 10 nm, likely due to difficulties in fitting the temperature profile at small scales.

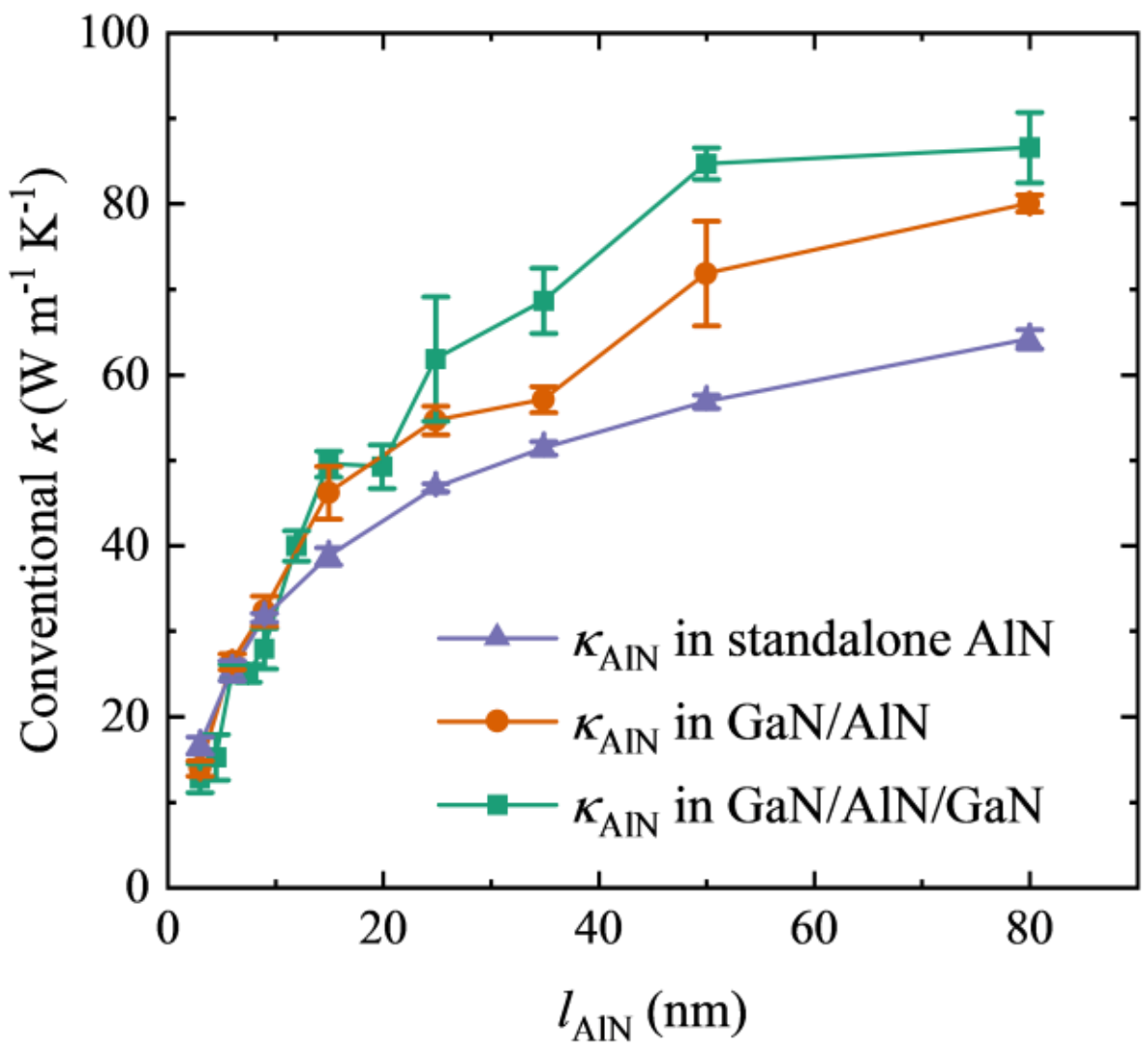

FIG. 6. Conventional thermal conductivity of AlN along the $c$ axis obtained from the MLIP-NEMD simulations. Quantum corrections are performed based on the ratio of quantum and classical specific heat of AlN.

## V. MULTIPLE INTERFACES

To further confirm that the TBC enhancement from nearby interfaces arises from the ballistic effects rather than coherent wave effects, we extend our analysis to multilayer structures. Here, the number of interfaces is varied from one up to five, as computational limitations prevent further increases with MLIP. For each multilayer configuration, we investigate two cases: one with equal layer thickness and the other with random layer thickness, while ensuring identical total thicknesses. For the random layer thickness cases, only one specific structure is examined, as demonstrated in the insets of Fig. 7. We do not expect major variation in the results across different combinations of layer thickness. In principle, coherent phonons are expected only in structures with equal layer thickness. The TBC of the first GaN/AlN interface ($G_1$) is measured in all cases, as shown in Fig. 7, where the schematic diagrams of the random-layer-thickness structures are also shown. A 6 nm layer thickness is used for equal-thickness structures. This is why the TBC of single interface case is slightly higher than 600 MW m$^{-2}$ K$^{-1}$, as explained in Fig. 3.

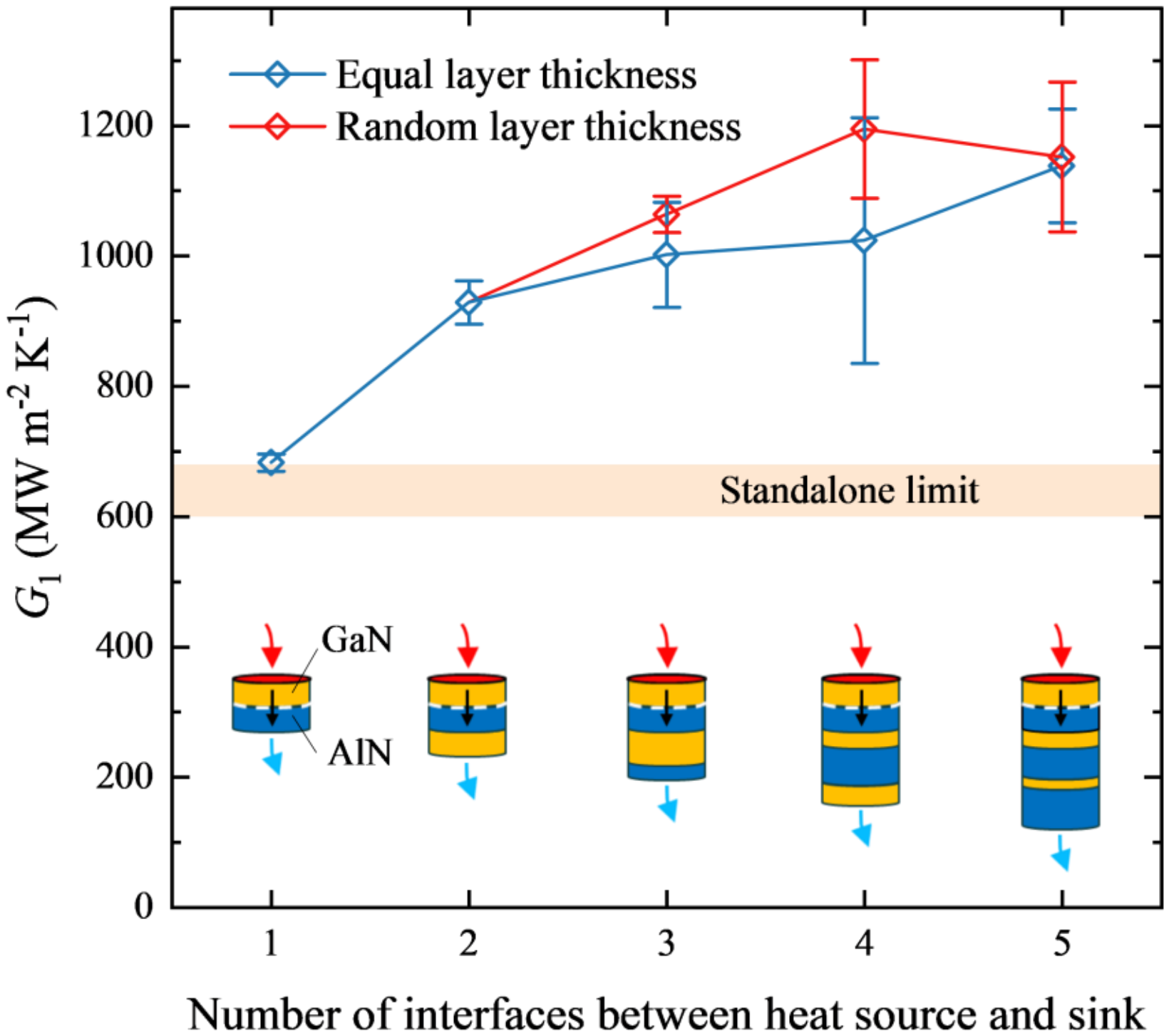

FIG. 7. The thermal boundary conductance of the first GaN/AlN interface as a function of the number of added layers. The insets show the configuration dimension of the random layer thickness cases.

As shown in Fig. 7, the original interface's TBC consistently increases with the addition of nearby interfaces, regardless of whether the layers have equal or random thicknesses. Since coherent effects would manifest only in the equal-thickness structures, this suggests that ballistic, rather than coherent, effects drive the observed enhancement. Interestingly, TBC is slightly higher for the random layer thicknesses than for equal layer thicknesses, which could be attributed to either the better ballistic effect due to some closer interfaces in random-layer thickness structures or the uncertainty. Additionally, more interfaces generally lead to higher TBC enhancement, likely because additional interfaces filter phonons more selectively. For instance, when a second interface is added, the TBC increases from 682 to 928 MW m$^{-2}$ K$^{-1}$; adding three more interfaces boosts it further to around 1150 MW m$^{-2}$ K$^{-1}$. Notably, this study focuses on assessing how nearby interfaces influence the TBC of the original interface rather than minimizing the total thermal resistance of the multilayer structure.

For comparison, Chen *et al*. studied GaN/AlN/GaN/AlN multilayers using BTE and DMM and found the TBC values of the three interfaces to be 952, 1162, and 840 MW $m^{-2}$ $K^{-1}$ [43], respectively, which are all higher than the standalone GaN/AlN TBC (~550 MW $m^{-2}$ $K^{-1}$, as determined by Hu *et al*. using the same method [26]). Additionally, Wu *et al*. investigated the total TBC of GaN/AlN superlattices using NEGF by varying the number of periods while keeping the total thickness constant [24]. It is known that usually the total TBC is reduced with more interfaces. However, they observed a decreasing total TBC followed by an increase as the number of interfaces grows, which may imply that the ballistic effect increases the TBC of one individual interface. However, superlattice structure is used in their study, which raises a question of how to decompose the contributions of ballistic and coherent effects. Further investigation is needed to answer this question. Moreover, though GaN/AlN multilayer or superlattices have been studied extensively [17,92], most work focuses on the overall thermal conductivity or TBC of the entire structure rather than the TBC at individual interfaces. Therefore, we do not compare our results with these works.

## VII. CONCLUSION

In conclusion, using MLIP-based NEMD, we determine that the theoretical TBC of a standalone GaN/AlN interface at room temperature is approximately 600 MW $m^{-2}$ $K^{-1}$, with negligible size effects. Comparisons with first-principles NEGF results suggest that inelastic transmission becomes significant above 100 K, reaching ~40% at 300 K. Additionally, we show that TBC can be significantly enhanced by the presence of a nearby interface or multiple nearby interfaces. This enhancement is due to the ballistic effects of the original material's bulk phonon mode, rather than the emergence of coherent superlattice phonon modes. Based on this behavior, we introduce a concept: "critical separation distance" $l_{\mathrm{cs}}$, representing the threshold distance beyond which the impact of nearby interfaces on TBC becomes negligible, and the interfaces become independent. This critical distance $l_{\mathrm{cs}}$ is longer than the MFP of the phonon modes that pass through the interfaces and transport inside the sandwiched material. This work addresses questions about thermal transport across GaN/AlN heterostructures and sheds light on the thermal management of electronic devices with nanoscale multilayer structures.

**Acknowledgments**

This work is partially supported by the National Science Foundation (NSF) (Award No. CBET 2337749). H.Z. and T.F. thank the support from the Nano & Material Technology Development Program through the National Research Foundation of Korea (NRF), funded by the Ministry of Science and ICT (RS-2024-00444574). The computation used Center for High Performance Computing (CHPC) at the University of Utah and the Advanced Cyberinfrastructure Coordination Ecosystem: Services & Support (ACCESS) program.

**Competing interests**

The authors declare no competing interests.